# Internet of Things(IoT) Based Multilevel Drunken Driving Detection and Prevention System Using Raspberry Pi 3

Viswanatha V[1], Venkata Siva Reddy R[2], Ashwini Kumari P[3], Pradeep Kumar S[4]
[1]Department of ECE, Acharya Institute of Technology, Bengaluru, India, viswas779@gmail.com
[2]School of ECE, REVA University, Bengaluru, India, venkatasivareddy@reva.edu.in
[3]School of EEE, REVA University, Bengaluru, India, ashwinikumari.p@reva.edu.in
[4] Department of TCE SIR MVIT, Bengaluru, India, Pradeep.appu@gmail.com

*Abstract —* **In this paper, the proposed system has demonstrated three ways of detecting alcohol level in the body of the car driver and prevent car driver from driving the vehicle by turning off the ignition system. It also sends messages to concerned people. In order to detect breath alcohol level MQ-3 sensor is included in this module along with a heartbeat sensor which can detect the heart beat rate of driver, facial recognition using webcam & MATLAB and a Wi-Fi module to send a message through the TCP/IP App, a Raspberry pi module to turn off the ignition and an alarm as prevention module. If a driver alcohol intake is more than the prescribed range, set by government the ignition will be made off provided either his heart beat abnormal or the driver is drowsy. In both the cases there will be a message sent to the App and from the App you can send it to family, friend, and well-wisher or nearest cop for the help. The system is developed considering the fact if driver is drunk and he needs a help, his friend can drive the car if he is not drunk. The safety of both the driver and the surroundings are aimed by this system and this aids in minimizing death cases by drunken driving and also burden on the cops.**

## I. INTRODUCTION

The rapid development of Indian economy and rise of standard of living of people leads to increase the number of vehicles and drivers. The increase of vehicles results in convenient life, economic profit and high quality of life for the people However it also brings traffic and road accidents as well. Nowadays, road accidents happen very frequently, which results in severe economic loss and people injury or death. Drunk driving, over speeding and fatigue are the three main reasons for the rise in road accidents, among which, drunken driving based road accidents are quite high in proportion. The point of execution of the framework is to diminish the tally of mishaps happening because of inebriated driving utilizing IOT.

Day by day, the usage of IOT (web of things) is expanding in the regions of digital security as well as luxury of human related issues. The implementation of IOT applications become simple due to easy of programming languages like python. Anybody with minimum programming skills can learn python. Size of the code is reduced drastically using python. Lot of libraries are available for python and Git hub is main source of code and libraries. After all these home automation become simple application using wireless protocol like Bluetooth. An individual sitting before TV can control different devices like fan, light, geyser and etc. Even the doorbell can automatically ring as soon as someone appears in front of main door and there is no one to attend the visitor.

The design of beautiful IOT application is possible just because of Raspberry Pi gadget which resembles a motherboard, with mounted chips and ports exposed .This way of architecture design simplified the availability of input ports, output ports and storage devices.

Raspberry pi gadgets are mostly characterized into two sorts; Model A and Model B. The fundamental contrasts are the expansion of USB port and Ethernet on the more costly model B. This paper uses the later one. Main devices used in the hardware implementation are as follows.

### A. Raspberry Pi board

Raspberry Pi 3 model B is a Master card estimated single-board PC (SBC) planned in the UK. Since its first discharge in 2012, numerous models of the Pi have been discharged from $5 to $35.More than 20 million units of the Pi were sold by 2018. This turns into the primary model to include remote availability. Wi-Fi 802.11n and Bluetooth 4.1.this can likewise boot from a USB. The Pi 3 model B+ and pi 3 model A+ are presented with 802.11ac and Bluetooth 4.2 in 2018 [8].There are different segments on the Raspberry Pi 3 board as appeared in fig.1.Its perfect for fledglings to find out about PCs and programming. Python and scratch stopped naturally alongside Raspbian as the OS. Web server can be arrangement on the Pi. The Pi can deal with playing sound/video, spilling video or be utilized as a media focus. Pi is additionally a decent stage to learn gadgets, interface different sensors to construct IOT ventures. The little structure factor and lower control necessities of Pi empowers battery controlled applications, generally in remote regions.

It is additionally utilized as an IOT door. Since the requirement for edge preparing and investigation being expanded, the Pi has been utilized for such purposes. All in all, the Pi is increasingly reasonable for applications that require part more information handling, coordinate with different gadgets, include media spilling and deal with various procedures while a microcontroller-based system(such as Arduino) is great at info/yield.

While a few models evacuate a portion of these interfaces, coming up next are available on Pi 3 Model B+:Gigabit Ethernet, simple sound/video 3.5mm jack,4x USB 2.0,HDMI,Camera sequential interface(CSI),Display sequential interface(DSI) and a 40-stick GPIO header pins can associate with a camera. DSI can associate with a touchscreen show. For remote systems administration, there's Bluetooth 4.2 BLE and 802.11 air conditioning. GPIO header pins underpins advanced interfaces: UART, SPI and I2C.There are 17 pins for GPIO. It is likewise having I2S interface for sound yield. All GPIO keep running on 3.3V and can give most extreme 20mA.Due to these highlights. The Pi is being utilized for IOT and mechanical technology applications. One interface that the Pi needs is analog to Digital converter (ADC).Therefore outside ADC must be utilized to interface to simple sensors. The Pi can be extended with HAT (Hardware Attached on Top) development sheets..
There have been to date three ages of Raspberry Pi. Gen1, Gen2 and Gen 3.The "in addition to" variations are superior to the plain ones of that age. Third era models are where remote availability was presented just because. Higher age sheets have more memory and quicker CPU. Regarding structure factor, ability and cost, in diminishing request of B, A and Zero models.

*B. Features of Raspberry Pi 3*

Various components available on Pi-3 module are as shown in fig.1 and also listed as follows.

- ARM CPU/GPU -- It's 64 bit system on chip (SoC) Broadcom chip "BCM2837" used in the design of architecture Pi 3. It is a four core ARM cortex A53 processor that implements the (ARMv8) architecture which run at 1.2GHz .This architecture is suitable for web browsing applications. It also has GPU Broadcom video core IV. This makes about 50% faster than the Pi 2.

- GPIO – These are general purpose input / output pins projected outside for interfacing external devices like sensors. There are 40 pins available on board.

- Audio out – It's a standard 3.55-millimeter sound jack utilized for sound yield through earphones or speakers. In any case, there is no sound in

- LEDs – These diodes, present on board to indicate the status of various processes happening over the board.

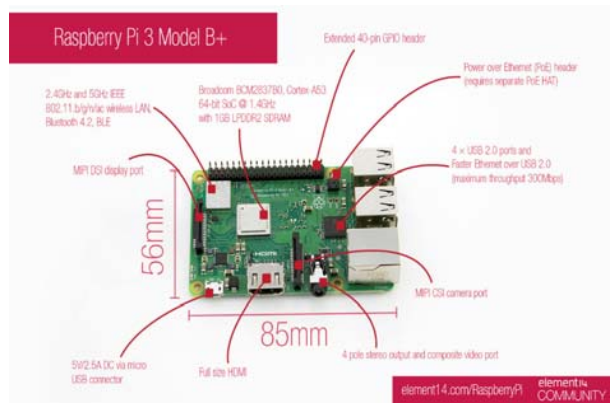

Courtesy :element 14 community
Fig.1 Raspberry Pi 3 module

- USB – Universal Synchronous bus, its input port that makes the connection for peripheral various devices such as mouse, keyboard, webcam etc... Also the power drawn by the Raspberry pi is through USB port. The recommended input voltage is 5V.for Pi 3, USB 2.0 ports are used.

- HDMI – This connector allows to use HDMI-compatible TV and desktop computer for audio and video processing. The original devices highlights HDMI and RCA however he ongoing model like Pi 3, has just the HDMI and the otherworldly 3.5 mm media port. This connector enables you to attach a top notch TV or other good gadget utilizing a HDMI link.

- Input power supply – It receives 5v from any compatible power supply through micro USB

- SD card slot – For installation of NOOBS or the image of rasbian, the minimum size of 8GB. Pi 3 uses SD card for storage of OS, libraries and user programs. Memory card of size 32GB is ideal for Pi 3.The best memory card is SanDisk Ultra 32GB.The Raspberry Pi .img or .zip file that is supposed to be loaded into memory card to be downloaded from internet and by using Etcher tool image file can be loaded into memory card.

- Ethernet – It's used for wired internet access. It's available on Pi 3 model B. But it doesn't support gigabit Ethernet. Only Raspberry Pi 3 model B+ does. onwards

- Wi-Fi –The Raspberry Pi 3 has come with an on board 802.11n wireless LAN adapter.so it's easy to set up Wi-Fi connection between internet source and Pi by log into the Pi with PuTTY in case OS Window is used or MiniCom in case OS Linux is used. Use SSID name which is name of Wi-Fi network given and password.

One can go about setting up a Raspberry pi3 as depicted in fig.2

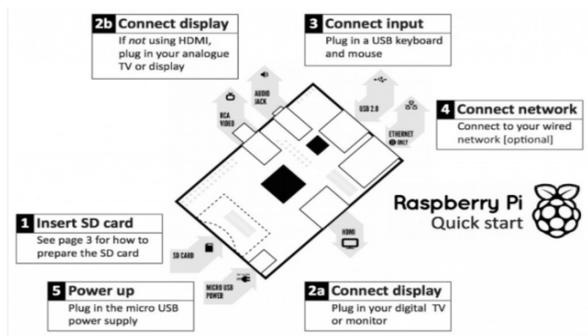

Courtesy: Joseph 2013

Fig.2 Steps to connect Raspberry Pi to peripherals.

The software support for Raspberry Pi that it runs with its own operating system (OS) so called real time OS (RTOS). The official OS bolstered by the establishment is Raspbian, which depends on Debian Linux dissemination. Ubuntu MATE, Snappy Ubuntu Core and RaspBSD are the other alternatives which are again Linux-based OSs. There also windows 10 IOT Core. Android Things, PiNet, Weather Station and Ichigojam Pi. To use Pi as a media, Open ELEC, LibrELEC, LibreELEC or RISC OS developed in the 1980s for ARM processors can be used. All these OS can be installed directly or using the NOOBS installer. Further software support rely on operating system use. For example, Rasbian comes pre-installed with Java, Python, scratch sonic pi and more. Python libraries Rpi. GPIO and Wiring Pi are useful for IOT programming.

### C. Literature Survey

A model for detection of alcohol using sensors such as MQ-3 and IOT are discussed in [1]. The Microcontroller STC12C516A is used as heart of the system and makes the decision based on input signals given by alcohol sensor. It finds the alcohol contentment in a person's breath [2]. Along with microcontroller, the system also contains LCD display, alcohol sensor, GPRS module. The system is placed in the vehicle and if the alcohol content is high than the limit, immediately the buzzer starts ringing. Level of alcohol content is displayed on LCD and at the same time the message will be sent to nearest authorities with particular location of the car and the car would be turned off automatically. Merits: a).Low cost & effective implementation. b).Alcohol sensor MQ-3 is very cheap. c). since it is a Microcontroller based model the size of the model is compact. d).When the system detect the threat, the car ignition automatically turns off. Disadvantages: a).Only a MQ-3 Sensor is used. b).The System threshold value is always a Static number.

the monitoring system with charge-coupled-device camera on real time basis to capture images of the person who is driving is discussed in [2]. Fatigue level of the driver is inferred by combining various visuals signals that are extracted in real time. Visual signals typically characterizes the level of activity of a person. The employed visual signals characterize gaze movement, movements in eyelid, head and facial appearance [3]. Human exhaustion is displayed utilizing a probabilistic demonstrating systems to anticipate weakness dependent on the got visual sign. The orderly blend of numerous visual sign yields a significantly more precise and hearty exhaustion portrayal than utilizing a solitary visual sign. The outcomes gotten by the framework was approved under genuine weakness conditions with human subjects of different genders, with/without glasses, different ethnic backgrounds and different illumination conditions. It was found to be reliable, accurate and reasonably robust in fatigue characterization.

Advantages: a).The proposed model is has high accuracy in detecting the threat. b).It has both Drunk & Drowsiness detection. Disadvantages: a).The size of the system is huge & wearing it makes the driver irritated. b). Wearing these sensors work, but to ensure that rider wears it is a problem.

The authors have designed the prototype which sends messages to their relatives for every five minutes [4]. The structured framework comprises of arm7 microcontroller, GSM and GPS modules. Position of the vehicle with longitude and scope are gotten by GPS module then GSM send same information to the relatives of the driver until he achieves home securely. The creators likewise utilized another element like mishap avoidance innovation utilizing ultrasonic sensor and GSM module. When accident happens, position of the vehicle will be sent continuously via GSM module. Advantages: a).No external checking is required. b).Hardware interface is only required. Disadvantages: a).Alcohol content of co-passenger also may get detected by gas sensor.

The authors in [5] suggested an efficient system aimed at early detection and prevention system. The system consists of Android G telephone with installed sensors like accelerometer and direction sensors. The product introduced on cell phone figures increasing speeds utilizing the information got from the sensors present in the cell phone and contrast it and run of the mill alcoholic driving examples taken from genuine driving tests. On the off chance that the information coordinates promptly, the telephone will naturally cautions the driver or call the police for assistance a long time before mishap really happens Advantages: a).The technique they have utilized is financially savvy. b). Procedure has no outer obstruction. Hindrances: Single calculation for the whole framework and all test conditions.

Over the years, several studies have been performed in the viewpoint of monitoring and detection of drivers. Moreover, many systems have been proposed and developed using different technologies as discussed in [6]-11].

### D. Working Principle

The Raspberry pi 3 model B system as shown in fig.3 is used as a controller to achieve the objective i.e. detecting the alcohol content and preventing the vehicle from accidents. Raspberry pi takes input from MQ-3 sensor which gives information about the alcohol content.MQ-3 sensor is a gas sensor which is sensitive to Alcohol, Ethanol, smoke and highly sensitive to alcohol and its sensitivity can be adjusted by the potentiometer [6]. It also takes input from heart beat sensor which gives information about the pulse rate of the driver which is fixed on the steering wheel. It takes input from webcam, which gives information about the drowsiness of the driver. Raspberry pi is interfaced to 16x2 LCD Display to display High or Low Alcohol and also to display pulse rate of the driver. It also interfaced to Wi-Fi module to get information of alcohol content and the pulse rate on Mobile phone app.MQ-3 sensor detects the Alcohol content in air. Wi-Fi is available to access the system for security and automation [7].Heart beat sensor detects the pulse rate of the driver. Wi-Fi is available to access the system for security and automation.

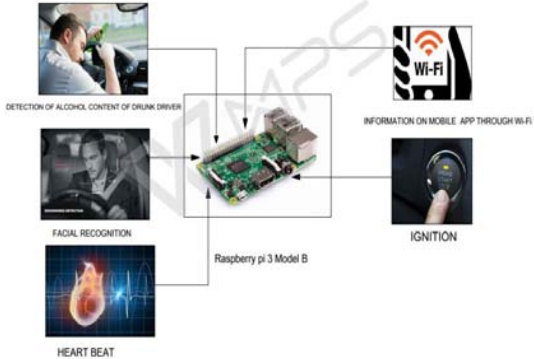

Fig.3.Block diagram of druken driving detection and prevention system

**Flow Chart**

Logic involved in monitoring and control of the system is as shown in the fig.4

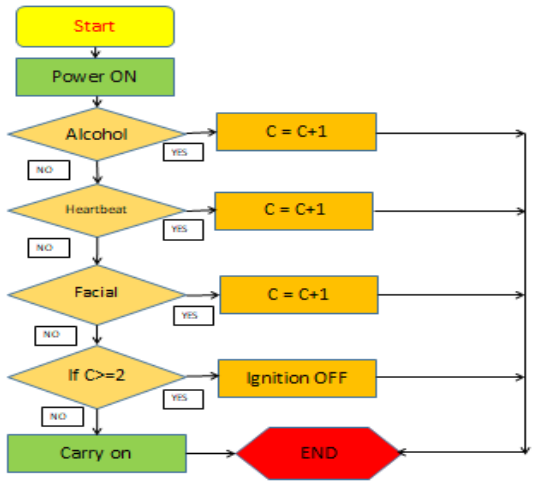

Fig.4. Flow chart

- When the driver sits on his couch the detection system will get activated.
- When the driver breathes the alcohol that's get into the air will be sensed by the MQ-3 sensor. If the alcohol content is above the legal limit the system counts '1', if not '0'.
- Simultaneously it moves to other two sensors i.e., heart beat sensor and facial recognition sensor.
- When heart beat is above 72beats/min then it counts '1', if not '0'.
- When driver becomes drowsy under the influence of alcohol then it counts '1', if not '0'.
- If the system matches any of the two sensors (C>=2) automatically ignition will be turned off.

## II. SOFTWARE TOOLS

Hardware of the above title has been implemented using Raspberry Pi-3.The following software tools are used in the process of hardware implementation.

- Raspbian: In Raspberry pi-3 board, Raspbian as an operating system which is basically Linux platform is used .This OS is installed in terms of image file on micro SD card via NOOBS software.
- Minicom: Its script based serial port communication tool .It is used to establish communication between external RS-232 devices such as mobile phones, routers and serial console ports. In this work, it is used to establish communication between programmer, Wi-Fi module, and Mobile phone to send the data to cloud. But before that the Raspberry Pi supposed to be configured and tested. This can be done using minicom program. This can be installed by running the command sudo apt-get install minicom.
- VNC (Virtual Network Computing) Viewer: Its software tool installed on Ubuntu Linux running on Laptop in order to access Raspberry pi remotely. To do so, make sure that both laptop and Raspberry pi connected to same Wi-Fi node.
- MQTT (Message Queuing Telemetry Transport): It is used to send data from sensors to the cloud where ThinkSpeak works as a server.
- ThinkSpeak is basically open source web server where data can be stored or retrieved from things by using HTTP protocol via local area network or over internet.It performs analysis and processing of data as it comes in over cloud.
- Matlab: (Matrix Laboratory): In the hardware implementation, this tool is used for Facial recognition which is a technology capable of finding or evaluating a person from a digital data represents image or a video frame from a video source. There are various algorithms used for facial recognition techniques but in general, they work by comparing selected facial features from given image with faces within a database.

## III. HARDWARE IMPLEMENTATION

The complete hardware implementation is pictorially drawn and as shown fig.5 as per the pictorial representation, the complete hardware is implemented and is as shown in fig.6. (a) and fig. 6. (b).

- Raspberry Pi consists of 40 GPIO (General purpose I/O) pins.

- Raspberry Pi takes input from MQ-3 sensor, heartbeat sensor and from the webcam ,process the data and the displays output on LCD display and also is sent to the mobile APP through Wi-Fi module(ESP12-R3)

- LED is used to represent the ignition system of the vehicle.

MQ-3 sensor: It consists of four pins such as Vcc, Digital output, Analog output, Ground. It can give both analog as well as digital output. Here only Dout is used because of ease as Raspberry Pi only accepts digital inputs.

- If Aout is used then use A/D converter (A/D converter compatible with Raspberry pi is MCP3008)

- Vcc pin of MQ-3 sensor is connected to PIN2 (common Vcc for purpose of ease) of Raspberry pi.

- Dout pin of MQ-3 sensor is connected to PIN15 (GPIO23) of Raspberry pi.

- GND f MQ-3 sensor is connected to PIN9 (common GND for purpose of ease) of Raspberry Pi.

Heart beat sensor: It has three pins i.e. Vout, Vcc (5v), GND.

UART: It is used for the purpose of serial communication. It is used to get the data from MATLAB code that is run for processing the data of facial recognition, Viola jones Algorithm is used to process the different images.

- It has three pins i.e. Rx, Tx and GND.

- Here Rx is not used because we are only transmitting the data to Raspberry pi but we are anything from it.

- Tx pin of UART is connected to pin10 (GPIO15) i.e.UART0_RXD pin of Raspberry pi.

- GND of UART is connected to pin 34 of Raspberry pi (Ground).

    LCD Display (16x2). It has 16 pins i.e, GND, Vcc, RS (Reset), R/W (read or write), EN (enable), D0-D7 and Led+ Led-.

- GND pin of LCD is bind to PIN1of Raspberry Pi.

- Vcc pin of LCD is bind to PIN2 (common Vcc) of Raspberry Pi.

- RS pin of LCD is to PIN40 (GPIO21) of Raspberry Pi.

- EN of LCD is given to PIN29 (GPIO5) of Raspberry Pi.

- D1, D5, D6 & D7 pins of LCD are bind to PIN31 (GPIO6), PIN33 (GPIO13), PIN35 (GPIO19) and PIN37 (GPIO26) respectively.

Wi-Fi module (ESP12-R3). It has 22 pins out which we are using only three pins i.e., Vcc, RXD0, GND and Wi-Fi network of this module should be connected to mobile phone to messages on mobile APP (TCP/IP app).

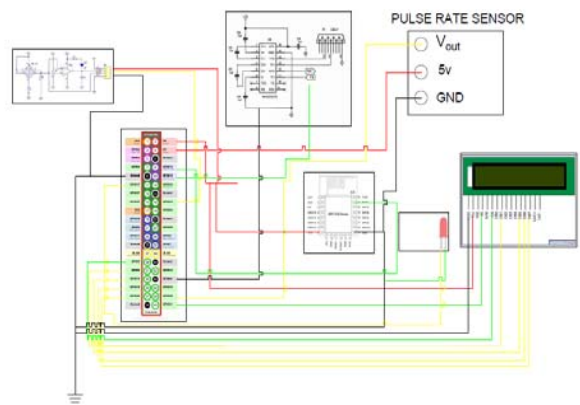

Fig.5.Complete circuit diagram.

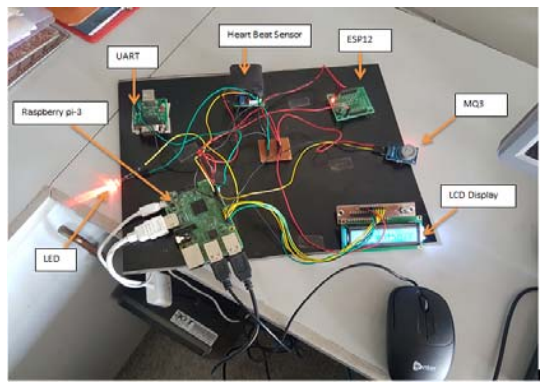

fig.6.(a)

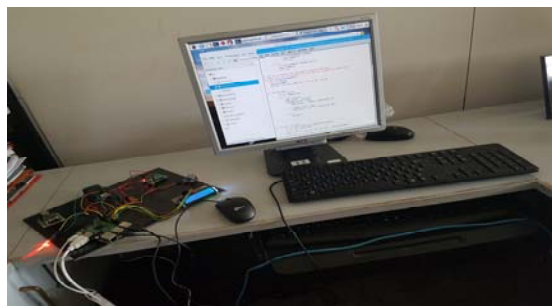

fig.6.(b)

Fig.6.(a) , 6(b) Complete Hardware Setup

### A. Alcohol Detection Sensor

An alcohol sensor as shown in fig.7 recognizes the alcohol present all around and generate equivalent voltage. Based on the level of voltage, the level of alcohol can be calculated in the human body which received alcohol[15]-[17]. MQ3 sensor works at the temperature of -10 to $50^0$ C with power supply of 5V and 150mV. The detecting range is from 0.04mg/L to 4mg/L which is suitable range for breath analyzers[12]. The sensitivity of MQ3 is high in alkali, sulfide, benzene steam, smoke and others.

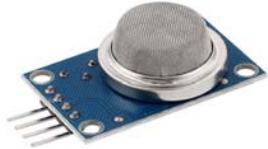

Fig.7. alcohol sensor

### B. Heart Beat Sensor

The heart beat sensor is of the new model TCRTI1000 has been used in the hardware to find out the pulse rate of the person who used alcohol and hence to calculate the beats per minute (BPM). If the value of BPM exceeds threshold value, it concludes that the person is drowse and the system will stop the vehicle from turning on. TCRT100 sensor basically optical sensor for photo plethysmography[13]. It consists of transmitter and receiver of photo-transistors which are arranged side-side in a leaded packages in order to block the surrounding ambient light. Otherwise it affects the sensor performance. This sensor module outputs analog as well as digital outs. Here we made use of digital output connected to Raspberry pi 3 which will further process the data received from heart beat sensor and calculate the BPM[14].

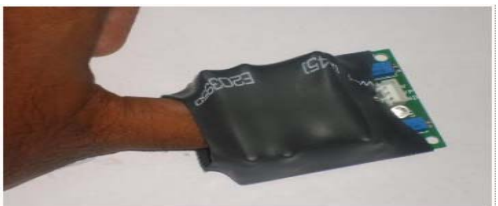

Fig.8. Heart beat sensor

### C. Detection of Face, Open Eyes Region

A Facial acknowledgment framework is novelty, well-equipped for distinguishing or confirming an individual using computerized picture or a video.

• There are certain strategies in which the frame works of the scanned images of the person do work. As a rule it compares with frames in database. If it matches, further action will be taken up by Raspberry Pi.

• An algorithm which is used analyze the relative size, position, shape of the jaw, nose, eyes and cheekbones. Later these features are used for comparison with frames in database.

• There are various algorithms exits in face recognition technology. Certain algorithms normalizes the bunch of face images followed by compression of data to minimize the usage of memory and improve the performance of the system which process the image. Later the probe image is compared with the test data.

• Low-resolution images of faces can be enhanced using face hallucination.

• A facial recognition is a technology which can verify or identify a person from digital image or video frame which can be obtained from video source.

• Again there are various methodologies to implement the technology of any kind. For example implementation of face recognition algorithms where nominated facial features gets compare with the faces in the records.

Results from the hardware setup for face recognition are as follows in two cases.

Case 1. Detection of face, Open Eyes region, when the driver is normal

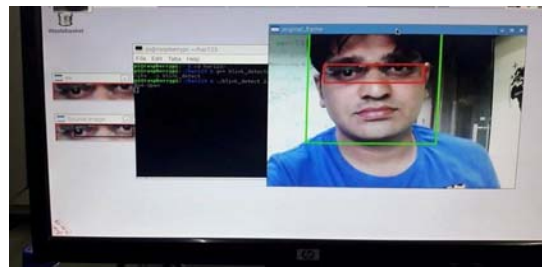

Fig.9. Detection of face when the driver is normal

Case.2 Detection of face, Closed Eyes region, when driver is drowsy.

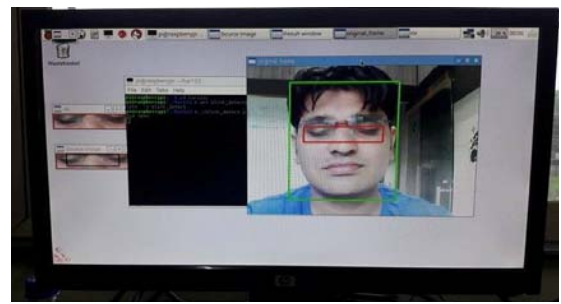

Fig.10. Detection of face when the driver is drowsy

## IV. CONCLUSION

In this paper, three levels of detection and control of drunken driving system is demonstrated through internet of things using Raspberry pi-3 as main control unit. The system automatically detects whether the driver is drunk or not and prevent him from driving the vehicle by turning off the ignition system. It also sends messages to concerned people.

One of the most robust level of detection and control used here is facial recognition technique which is an eye based control. In future, control of all types of devices will be based on Eye based control, thus control operation becomes much easier and comfortable with less human presence. Many operations which are risky can be performed easily by this kind of application. Further study and research on these technologies results in creation of new trend of interacting with machine so called machine learning.